\documentclass[useAMS, usenatbib]{mn2e}
\usepackage{graphicx}
\usepackage{amsmath}
\usepackage{amssymb}
\usepackage[T1]{fontenc}
\usepackage{aecompl}
\def\be{\begin{equation}}
\def\ee{\end{equation}}
\def\bea{\begin{eqnarray}}
\def\eea{\end{eqnarray}}

\title[Fractal basins of escape and the formation of spiral arms in
a galactic potential with a bar]{Fractal basins of escape and the formation of spiral arms in
a galactic potential with a bar}
\author[Andreas Ernst  and Thomas Peters]{Andreas Ernst$^{1}$\thanks{email: aernst@ari.uni-heidelberg.de} and Thomas Peters$^2$\thanks{email: tpeters@physik.uzh.ch}  \\
$^{1}$Astronomisches Rechen-Institut/Zentrum f\"ur Astronomie der Universit\"at Heidelberg, 
M\"onchhofstrasse 12-14,
69120 Heidelberg, Germany \\
$^{2}$Institut f\"ur Computergest\"utzte Wissenschaften, 
Universit\"at Z\"urich, Winterthurerstrasse 190, CH-8057 Z\"urich, Switzerland
}

\begin{document}

\date{Accepted ... Received ...}

\pagerange{\pageref{firstpage}--\pageref{lastpage}} \pubyear{2002}

\maketitle

\label{firstpage}

\begin{abstract}
We investigate the dynamics in the close vicinity of and within the critical area
in a 2D effective galactic potential with a bar of Zotos.
We have calculated Poincar\'e surfaces of section
and the basins of escape. 
In both the Poincar\'e surfaces of section and the basins of escape 
we find numerical evidence for the existence of a separatrix which 
hinders orbits from escaping out of the bar region.  
We present numerical evidence for the similarity between spiral arms of barred spiral 
galaxies and tidal tails of star clusters.
\end{abstract}


\raggedbottom

\section{Introduction}

Many spiral galaxies have barred central regions. 
It has been established by \citet{deVaucouleurs1963} that roughly one third of disk galaxies are strongly 
barred, one third do not have a bar, and the remaining third are, with respect to having the bar property, 
of intermediate or undeterminable type.

There was a long controversy whether the bar fraction of barred to disk galaxies is 
red-shift-dependent. Recent results suggest indeed that the fraction of barred spirals declines 
with redshift \citep{Sheth2008, Masters2011, Melvin2013}.

The stellar-dynamical reason for the occurence of bars is believed to be a dynamical instability in rotationally
supported stellar discs \citep{Miller1970,Hohl1971,Ostriker1973,Sellwood1980}.
In particular, if the ratio of rotational to random kinetic energies exceeds a certain threshold,
an initially axisymmetric stellar system is unstable to the formation of a bar-like mode, i.e. a 
non-axisymmetric perturbation.

The present study aims at numerically investigating the dynamics in the vicinity of a bar in
the centre of a galaxy. In the language of dynamical systems theory, the issue amounts
to examining bound and unbound orbits in a two-dimensional Hamiltonian system. The problem of escape
from Hamiltonian systems is a classical problem in dynamical astronomy
and nonlinear dynamics \citep[e.g.][]{Contopoulos1990,Contopoulos1992,Contopoulos1993,Contopoulos2012,
Siopis1996,Navarro2001,Schneider2002,Ernst2008}. However, it is far less studied than the closely
related situation of chaotic scattering, where a body from infinity approaches and scatters off
a complex potential. This problem is well understood from the viewpoint of chaos theory
\citep[e.g.][]{Bleher1989,Bleher1990,Boyd1992,Boyd1993,Chen1990,Ding1990,Eckhardt1986,Eckhardt1987,
Eckhardt1988,Gaspard1989,Henon1988,Jose1992,Jung1987a,Jung1987b,Jung1989,Jung1990,Jung1991,Jung1995,Jung1999,
Lai1993,Lai2000,Lau1991,Lipp1999,Motter2002,Ruckerl1994,Seoane2006,Seoane2007,Seoane2008,Sweet2000}
and has been applied in the astrophysical context to, e.g., the scattering off black holes
\citep[e.g.][]{Aguirregabiria1997,deMoura2000}
and three-body systems \citep[e.g.][]{Hut1983a,Hut1983b,Benet1999,Benet1996}.

We here do not study in detail individual orbits of stars within the galactic potential and Poincar\'e sections
\citep[e.g.][]{Athanassoula1983,Caranicolas1998,Contopoulos1980,Contopoulos1983a,Contopoulos1983b,Contopoulos1987,Hasan1990,Hasan1993,
Henon1964,Martinet1990,Pfenniger1994,Teuben1985,
Zotos11,Zotos2012,Zotos12b,Zotos12c},
but rather focus on the computation of the basins of escape, the related invariant
manifolds of the chaotic saddle associated with the chaotic dynamical behaviour \citep[e.g.][]{Ott2002}
and the formation of spiral structure as a result of the escape process.

The basins of escape are defined as those initial conditions (e.g., on a surface of section)
for which particles escape through exits in the equipotential surfaces around the Lagrangian points $L_1$ and $L_2$.
These exits open up for Jacobi energies which are higher than the critical Jacobi energy.
The critical Jacobi energy is defined to be the effective potential at 
the Lagrangian points $L_1$ and $L_2$.
The boundaries between the basins of escape may be fractal \citep{Bleher1988} or, as is the case for 
the widely known H\'enon-Heiles system \citep{Henon1964}, respect the more 
restrictive property of Wada \citep{Aguirre2001} in the case of three or more coexisting basins of escape.

The stable (or unstable) manifolds of the chaotic saddle are defined as the set 
of points on the boundaries between the basins of escape, for which orbits do not escape 
for $t\rightarrow\infty$ (or $t\rightarrow-\infty$). 
The chaotic saddle is defined as the intersection of its 
stable and unstable manifolds. 
The intersection points between the 
stable and unstable manifolds of the chaotic saddle are 
also called biasymptotic points \citep[c.f.][]{Simo2014}.
Both hyperbolic and non-hyperbolic chaotic saddles occur in dynamical systems \cite[e.g.][]{Lai1993}. 
A non-hyperbolic chaotic saddle may display vanishing splitting angles between the
stable and unstable manifolds \cite[][]{Lai1993}. This phenomenon is called a tangency.

In this context, it is worth mentioning that invariant manifolds of the Lyapunov orbits
\citep{Lyapunov1949} around the Lagrangian points $L_1$ and $L_2$ 
have been invoked recently
to explain the formation of rings and spirals in barred galaxies \citep{RomeroGomez2006,RomeroGomez2007,
Athanassoula2009,Athanassoula2009b,Athanassoula2010,Athanassoula2011,Athanassoula2012}.
These consist of
orbits which approach the unstable periodic Lyapunov orbits around $L_1/L_2$ asymptotically for $t\rightarrow +\infty$ or $t\rightarrow -\infty$.
A few of these asymptotic orbits are shown in Figure 8 of \citet{Fukushige2000}.

We refer to \citet{Seoane2013}, \citet{Aguirre2009} and \citet{Altmann2013} for recent reviews
on chaotic scattering, fractal basins and escape from chaotic systems, respectively.

This paper is organised as follows. In Section 2 we present the theory. In section 3 we discuss the results: Poincar\'e surfaces of section (section 3.1), orbits (section 3.2), basins of escape (section 3.3) and spiral arms (section 3.4). 
We summarise our conclusions in Section 4.

\section{Theory}

\begin{figure}
\includegraphics[angle=90,width=0.45\textwidth]{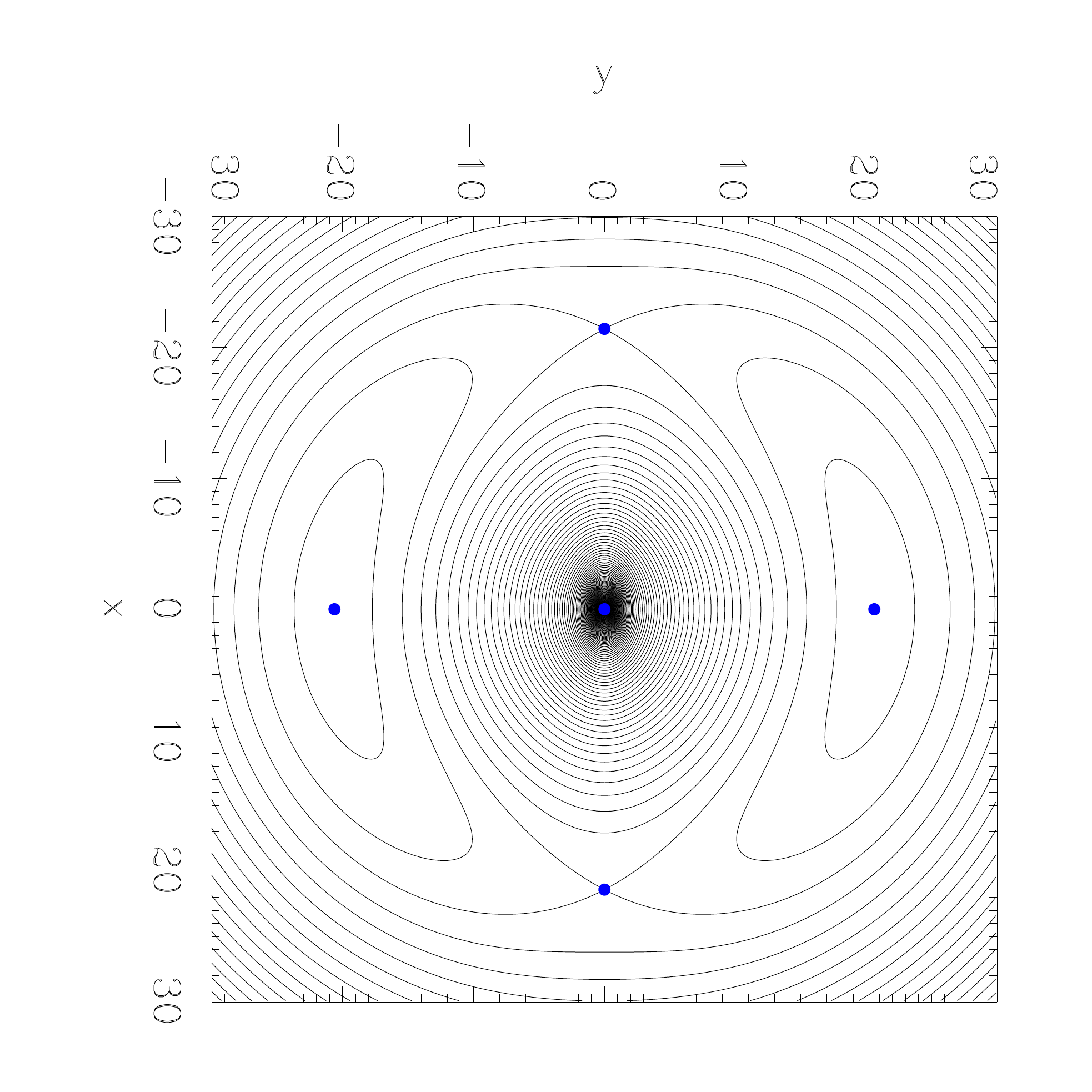} 
\caption{The effective potential of Eqn. (\ref{eq:zotospotential}) in 2D. The contours are the
isolines of constant effective potential. The Lagrangian equilibrium
points are visualized with (blue) dots.} 
\label{fig:effpotbar2}
\end{figure}

\begin{figure}
\includegraphics[width=0.5\textwidth]{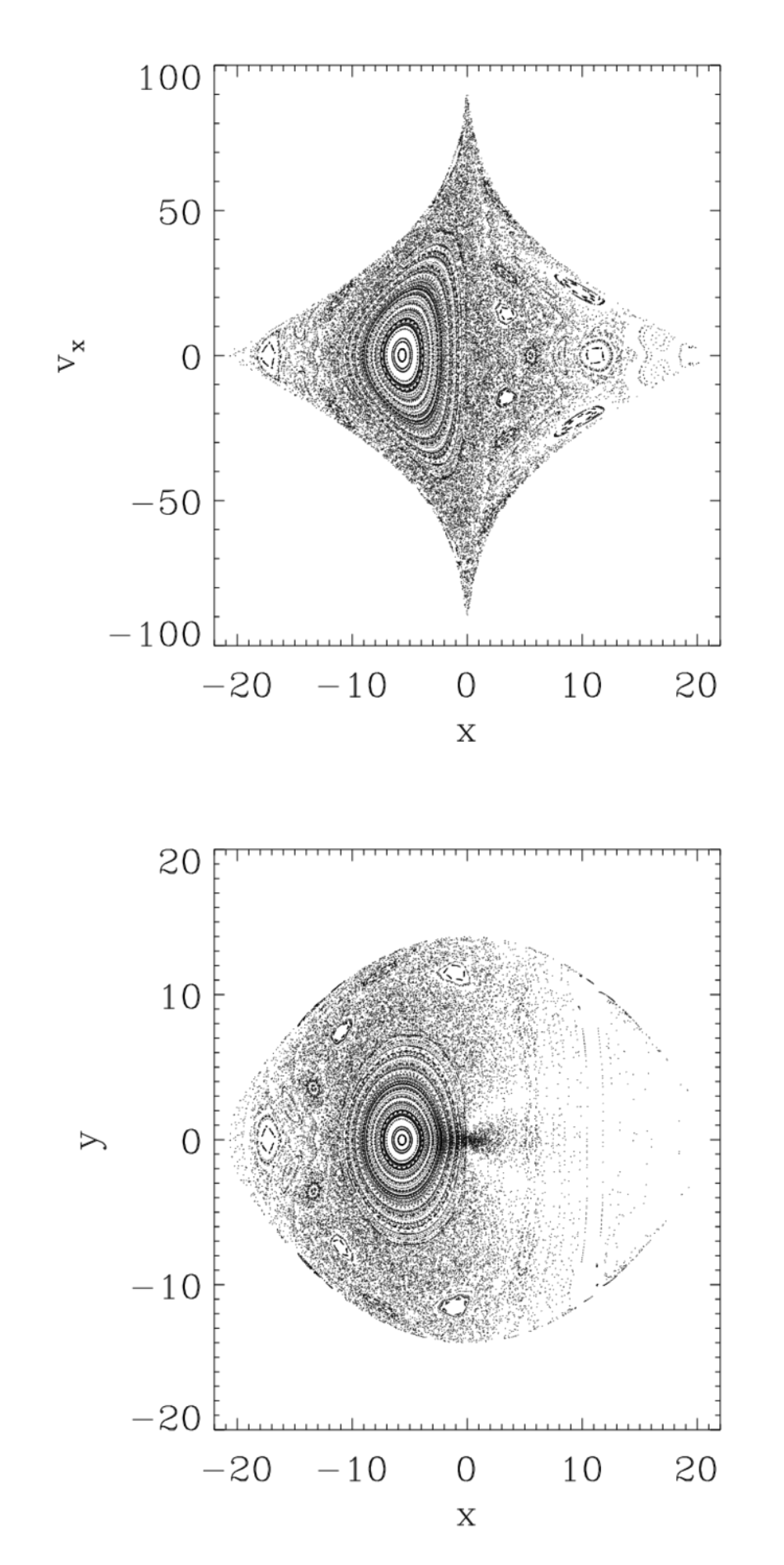} 
\caption{Poincar\'e surfaces of section at $e_{J,\rm crit}=-226.391170544787810$. 
Top panel: Crossing $y=0$ with $\dot{y}\leq 0$.
Bottom panel: Crossing $\dot{x}=0$ with $\dot{y}\leq 0$. 
} 
\label{fig:pbcrit}
\end{figure}

\begin{figure*}
\includegraphics[height=0.9\textheight]{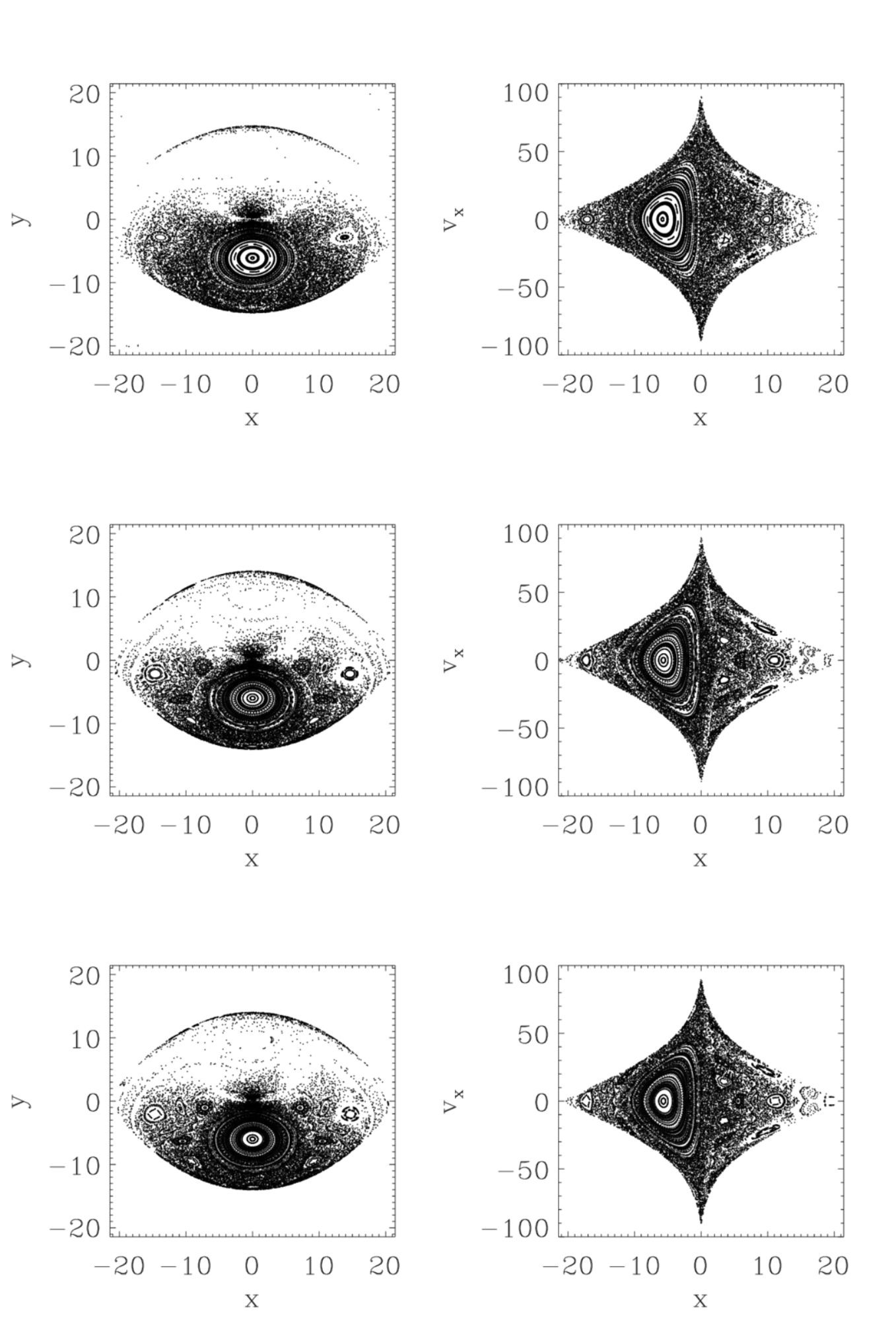} 
\caption{Poincar\'e surfaces of section 
at $e_J =-203.75205349030903$ (top panel), $-224.12725883933993$ (middle panel), $-226.16477937424302$ (bottom panel). Left panels: Crossing $\dot{y}=0$ with $\dot{x}\geq 0$. 
Right panels: Crossing $y=0$ with $\dot{y}\leq 0$.} 
\label{fig:pb2}
\end{figure*}

\begin{figure}
\includegraphics[width=0.4\textwidth]{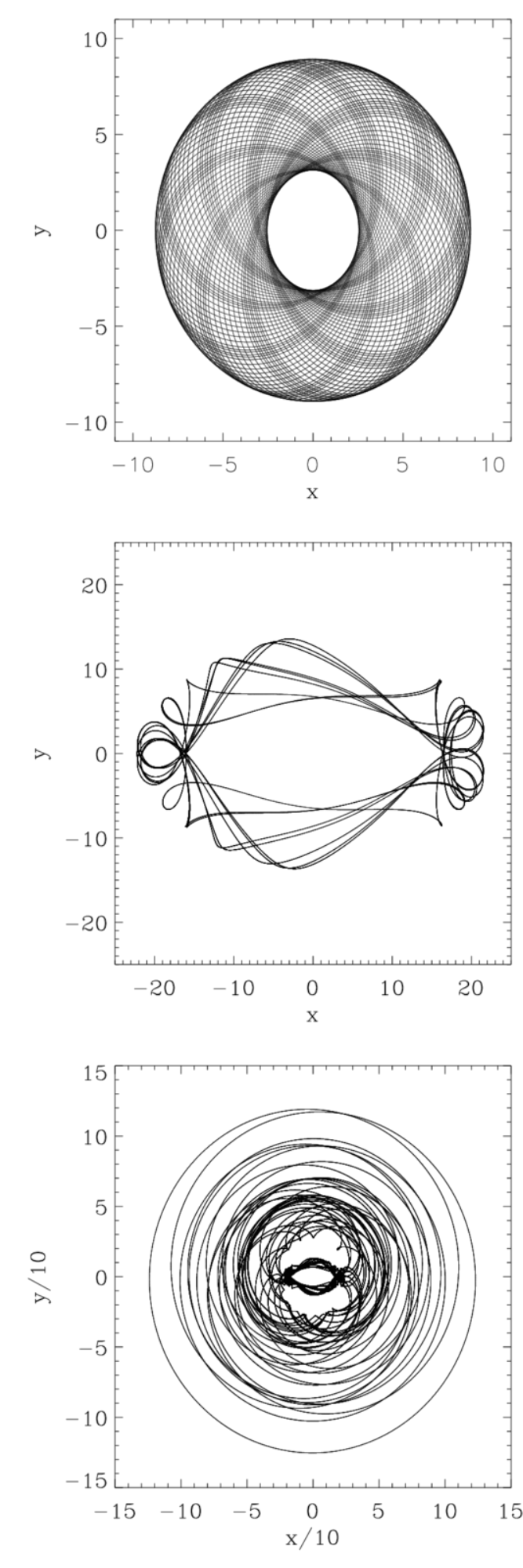} 
\caption{Four orbits. Top panel: Non-escaping retrograde quasiperiodic rosette orbit at 
$e_{J, \rm crit}$ with ($x_0,y_0,v_{x0})=(-3,4,0)$. 
Middle panel: Non-escaping chaotic orbit at $e_J =-203.75205349030903$ with ($x_0,y_0,v_{x0})=(-16,0,0)$.
Bottom panel: Escaping chaotic orbit at $e_J=-210$ with ($x_0,y_0,v_{x0})=(20.8,0,5)$. 
} 
\label{fig:orbits2d}
\end{figure}

We adopt the effective potential of \citet{Zotos2012}, which is visualised in Figure \ref{fig:effpotbar2},

\bea
\Phi_{\rm eff}(x,y) &=& -\frac{M_d}{\sqrt{x^2+y^2+\alpha^2}} -\frac{M_b}{\sqrt{x^2+b^2 y^2+c_b^2}} \nonumber \\
&&-\frac{M_n}{\sqrt{x^2+y^2+c_n^2}} + \frac{v_0^2}{2}\ln\left(x^2+\beta y^2 + c_h^2\right)   \nonumber \\
&&- \frac{1}{2}\Omega_b^2(x^2+y^2) \label{eq:zotospotential}
\eea

\noindent
with the parameters $\alpha = 8$, $\beta = 1.3$, $b = 2$, $v_0 = 15$, $M_d = 9500$, $M_b = 3000$, $M_n = 400$, $c_b = 1.5$, 
$c_n = 0.25$ and $c_h = 8.5$. We have $\Omega_b = 1.25$. The model consists of a disc, 
a bar, a Plummer nucleus (bulge) and a logarithmic halo. The bar rotates clockwise at constant angular velocity $\Omega_b$. The effective potential
in Eqn. (\ref{eq:zotospotential}) is the cut through the equatorial plane of a typical 3D potential of a barred
spiral galaxy. The parameters for all calculations in the present paper are chosen as in \citet{Zotos2012}. 
While the parameter $b$ determines the elongation of the bar,
$\beta$ determines the elongation of the halo. The weak effect of removing the non-axisymmetry in the halo by setting 
$\beta=1.0$ is shown as follows: The positions of the Lagrangian points $L_4$ and $L_5$ within the banana-shaped isolines of the effective potential in Figure \ref{fig:effpotbar2} are $(x,y)=(0,\pm 20.619528162205612)$ for $\beta=1.3$
and $(x,y)=(0,\pm 20.532965314242505)$ for $\beta=1.0$. The relative difference is $\Delta y/y=4.2\times 10^{-3}$, which is tiny.

In  Eqn. (\ref{eq:zotospotential}), the model units of length $L_0$, velocity $V_0$, angular velocity $\Omega_0$,
time $T_0$ and mass $M_0$ of the parameters,
in which the gravitational constant $G=1$, can be scaled to physical units of a barred spiral galaxy with the size of a galaxy such as NGC 1300 as follows:

\bea
L_0 &=& 1 \ \mathrm{kpc}, \ \ \ \ \ \ \ \ \ \ \ \ \  \label{eq:units1} \ M_0 \approx 2.223\times 10^7 \ M_\odot, \\
\Omega_0 &=& 10 \ \mathrm{km/s/kpc}, \ \ \ \ \ \ V_0 = 10 \ \mathrm{km/s}, \label{eq:units2} \\
T_0 &=& \sqrt{L_0^3/(G M_0)}\approx 100 \ \mathrm{Myr}. \label{eq:units3}
\eea

\noindent
In these physical units, the circular speed in the halo is $v_0=150$ km/s, the mass of the disk is
$M_d \approx 2\times 10^{11} M_\odot$ and the length of the bar is $2r_L \approx 42$ kpc.

The equations of motion in the rotating frame are given by

\bea
\ddot{\mathbf{r}} 
&=&  -\nabla \Phi_{\rm eff} - 2\left(\Omega_b\times\dot{\mathbf{r}}\right) \\
&=& -\nabla \Phi - 2\left(\Omega_b\times\dot{\mathbf{r}}\right) - \Omega_b \times \left(\Omega_b \times \mathbf{r}\right) \label{eq:eom}
\eea

\noindent
where $\mathbf{r}=(x,y,z)$ is the position vector, the dot denotes a derivative with respect to time 
and $\Phi$ is given by the first four terms in  (\ref{eq:zotospotential})
(i.e., without the last centrifugal potential term). Since we are considering a 2D case we set $z=0$
or neglect it.

The Lagrangian points $L_1$ and $L_2$ are defined by the condition

\be
\ddot{\mathbf{r}} = -\nabla_{\rm 2D} \Phi_{\rm eff} = \left(\begin{array}{c} 
0 \\
0 
\end{array}\right) \ee

\noindent
where $\mathbf{r}=(x,y), \nabla_{\rm 2D} = (\partial /\partial x, \partial /\partial y)$
and by the fact that they are saddle points of the effective potential, i.e. the two 
eigenvalues of the corresponding $2\times2$ Jacobi matrix at $L_1$ and $L_2$ are 
real and differ in sign.
In the case of the effective potential of Eqn. (\ref{eq:zotospotential}) 
$L_1$ and $L_2$ are located at $(x,y)=(0,\pm r_L)=(0,\pm 21.417693579040430)$.
$r_L$ is called Lagrangian radius. Let $L_1$ be at $x=-r_L$ and $L_2$ be at $x=+r_L$.
In Figure \ref{fig:effpotbar2}, there is a local minimum of the effective potential 
at $(x,y)=(0,0)$, which is usually called $L_3$, and two local maxima ($L_4$ and $L_5$) enclosed
by the banana-shaped isolines of the effective potential.

In the 3D case, the last closed equipotential
surface through $L_1$ and $L_2$ encloses the critical volume (in the 2D case it reduces to a critical area). 
In the context of star cluster, binary star or planetary dynamics, respectively, it is sometimes called Jacobi, 
Roche or Hill volume, respectively.
The Jacobi energy is an isolating integral of motion and defined by 

\be
e_J=\frac{\dot{\mathbf{r}}^2}{2} +\Phi_{\rm eff}
\ee

\noindent
The critical Jacobi energy is given by $e_{J,\rm crit}=\Phi_{\rm eff}(r_L) = -226.39117054478781$.
It is the effective potential of Eqn. (\ref{eq:zotospotential}) 
evaluated at the Lagrange points $L_1$ and $L_2$.
Orbits with $e_J < e_{J,\rm crit}$ are bound 
and cannot escape since there are no
exits in the equipotential surfaces around $L_1$ and $L_2$. 
If one increases the Jacobi energy for  $e_J > e_{J,\rm crit}$
the exits around $L_1$ and $L_2$ in the equipotential surfaces become larger.
We examine only situations with $e_J \ge e_{J,\rm crit}$, since we study the escape process
from the bar region.

We use in the present paper an eighth-order Runge-Kutta method for the orbit integrations.

\section{Results}

\subsection{Poincar\'e surfaces of section}

\label{sec:surf}

The Poincar\'e surfaces of section are 2D cuts through the 4D phase space. For example, for an
$x-v_x$ surface of section as in the top panel of Figure \ref{fig:pbcrit} we take an 
initial condition $x=x_0, v_x = v_{x,0}, y = y_0 = 0$ and choose 
$v_y$ positively as $v_y=\sqrt{2e_J - v_x^2 - 2\Phi_{\rm eff}(x_0,y_0)}$. 
Then we integrate the initial
condition forwards in time and plot a dot at each consequent (=``piercing point through the 
surface of section'') with $y=0$ when $v_y \leq 0$.

Figure \ref{fig:pbcrit} shows two Poincar\'e surfaces of section at the critical Jacobi energy. The top panel
shows orbits crossing $y=0$ with $\dot{y} \leq 0$. The bottom panel shows orbits crossing $\dot{x}=0$ with $\dot{y}\leq 0$.  

Figure \ref{fig:pb2} shows Poincar\'e surfaces of section at $e_J =-203.75205349030903$ (top panel), 
$-224.12725883933993$ (middle panel), $-226.16477937424302$ (bottom panel). 
These values of the Jacobi energy correspond to the relative deviations from the 
critical Jacobi energy $(e_{J, \rm crit}-e_J)/e_{J, \rm crit} = 0.1, 0.01$ and $0.001$ 
as in \citet{Ernst2008}. 

The following aspects are similarly found in the paradigmatic H\'enon-Heiles system \citep{Henon1964}:
\begin{enumerate}
\item In some regions an adelphic integral of motion is present which 
hinders the particles  on quasiperiodic orbits from escaping. 
\item In Figure \ref{fig:pb2}, it can be seen that with growing $e_J$, regular islands in the
Poincar\'e surfaces of section disappear and are replaced with regions that
show a chaotic dynamical behaviour.
\item Some areas on the surfaces
of section in the ``chaotic sea'' are less densely occupied by chaotic stellar orbits than others.
\end{enumerate} 

Regarding aspect (iii), it must be noted that, for $e_J > e_{J, \rm crit}$, 
the reason may be the fact that particles can leak out through exits in the equipotential surfaces
\citep[as in the upper right panel of Figure 2 of][]{Ernst2008}.

However, Figures \ref{fig:bs2} and \ref{fig:tbar} below confirm that a large subset of particles, which
includes even the chaotic orbits, cannot leak out at all and is trapped within a separatrix. 
In this case the fact that some regions are less densely populated by stellar orbits 
must have a different explanation. We found such a phenomenon also in our previous work
\citep{Ernst2008}.


We have also verified that that the Poincar\'e surfaces of section at $e_J = -570$ and $e_J=-2700$ are consistent
with Figures 2 and 4 in  \citet{Zotos2012}.

\subsection{Orbits}

In total, Figure \ref{fig:orbits2d} shows three orbits. 
The top panel shows a non-escaping retrograde quasiperiodic rosette orbit at 
($x_0,y_0,v_{x0})\approx(-3,4,0)$ at $e_{\rm J,crit}$.
The middle panel shows a non-escaping chaotic orbit at $e_J =-203.75205349030903$ with ($x_0,y_0,v_{x0})=(-16,0,0)$.
The bottom panel shows a typical example of an escaping chaotic 
orbit at $e_J=-210$ with ($x_0,y_0,v_{x0})=(20.8,0,5)$. 

\subsection{Basins of escape}

\begin{figure*}
\begin{center}
\includegraphics[height=0.9\textheight]{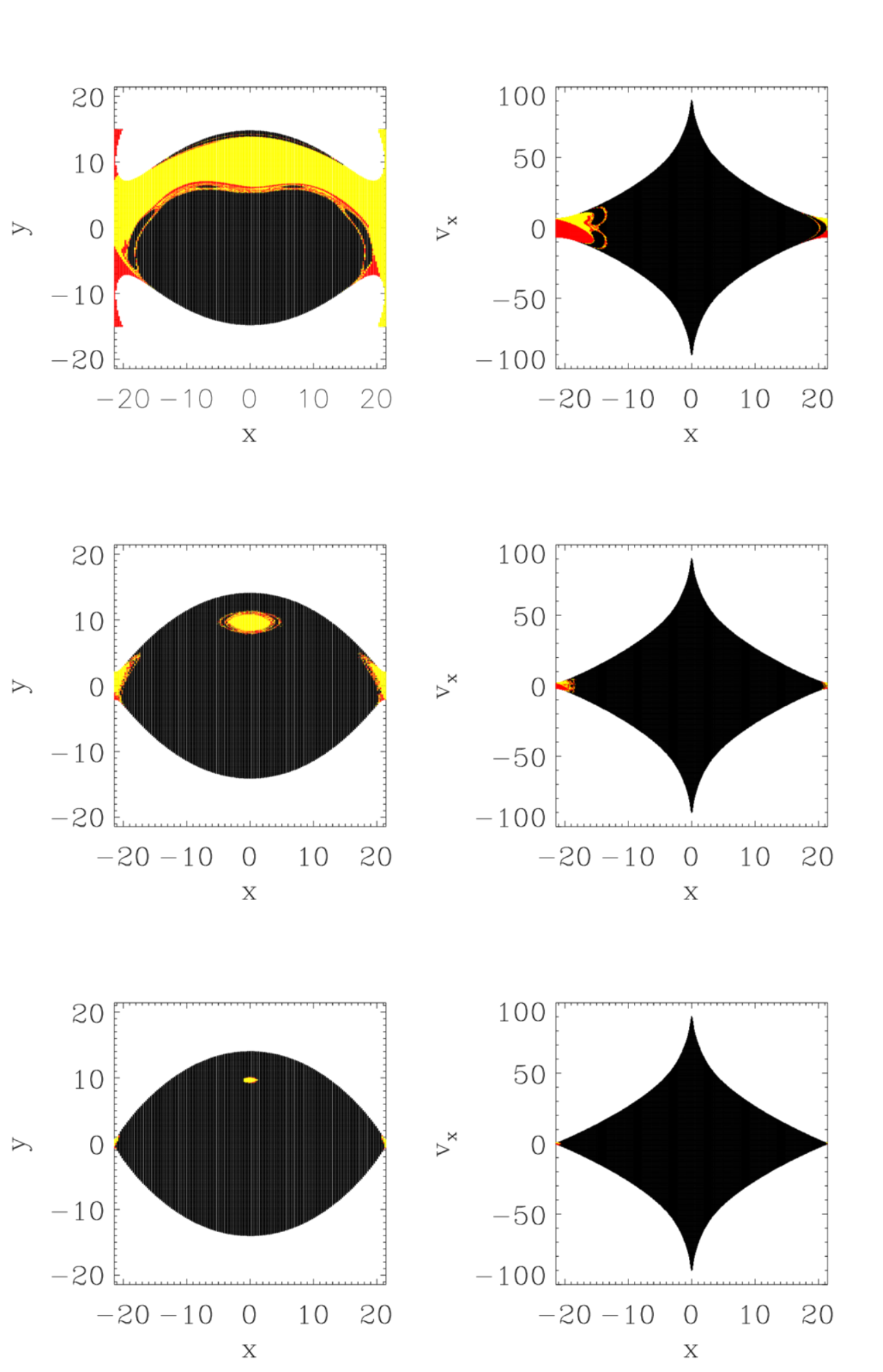} 
\caption{Basins of escape at $e_J =-203.75205349030903$ (top panel), $-224.12725883933993$ (middle panel), $-226.16477937424302$ (bottom panel).
The red (and yellow) regions correspond to initial conditions for which the escaping
star passes $L_1$ (or $L_2$). The black regions correspond to initial conditions for which
the orbit does not escape.} 
\label{fig:bs2}
\end{center}
\end{figure*}

\begin{figure}
\includegraphics[height=0.4\textheight]{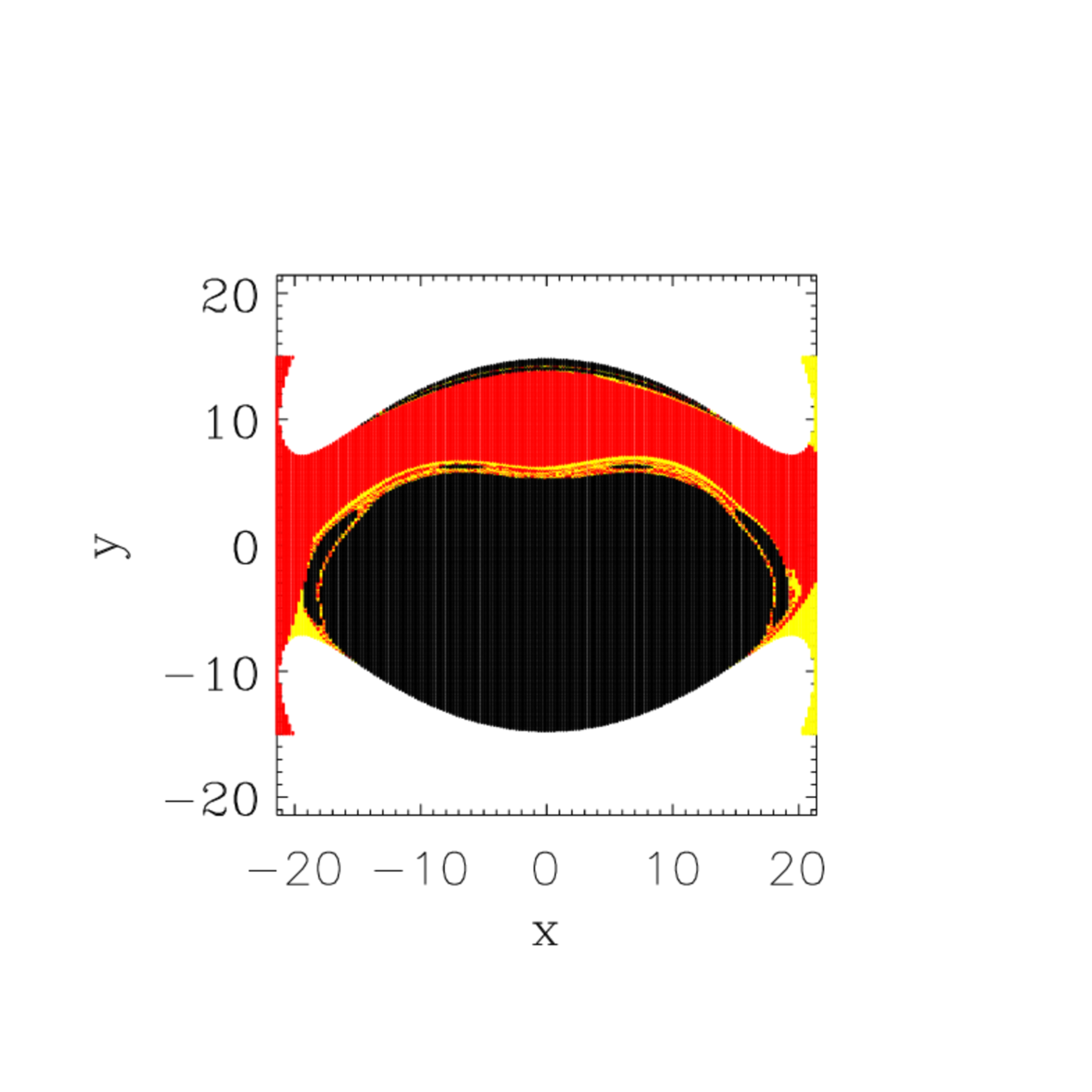} 
\caption{Backwards integration at $e_J =-203.75205349030903$. The red (and yellow) regions correspond to 
initial conditions for which the escaping
star passes $L_1$ (or $L_2$). The black regions correspond to initial conditions for which
the orbit does not escape.}
\label{fig:backwards}
\end{figure}

\begin{figure*}
\includegraphics[height=0.35\textheight]{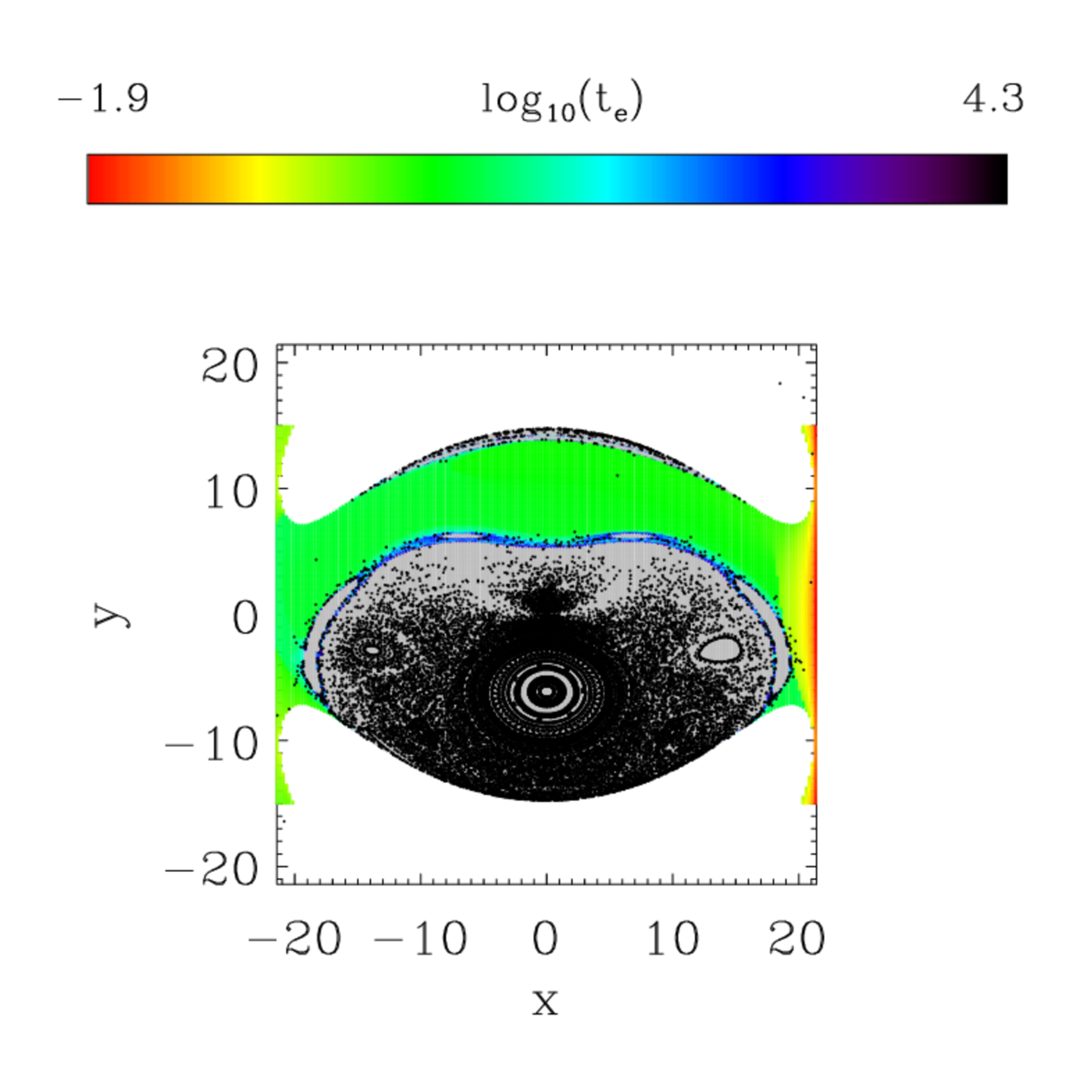} 
\includegraphics[height=0.35\textheight]{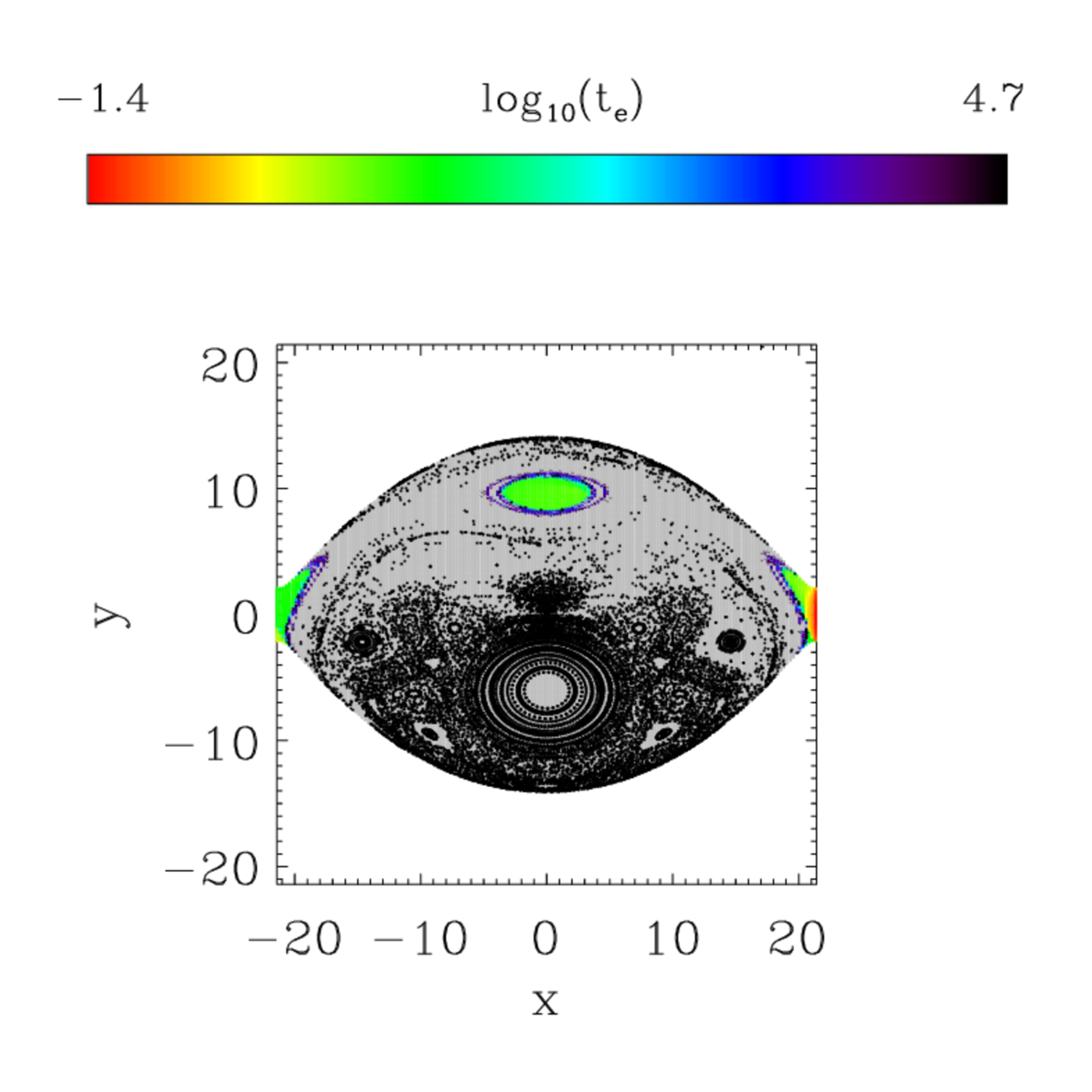} 
\caption{Distributions of escape times on surfaces of section. Left panel: 
At $e_J =-203.75205349030903$. Right panel: 
At $e_J =-224.12725883933993$. The scale on the color bar is logarithmic. The region within the separatrix,
from which orbits do not escape,
is overplotted with black dots and 
invariant curves on the grey-shaded area.} 
\label{fig:tbar}
\end{figure*}

Figure \ref{fig:bs2} shows the basins of escape for the same parameters as in
Figure \ref{fig:pb2}, i.e. for $e_J > e_{J, \rm crit}$.

For each panel on the left-hand side of Figure \ref{fig:bs2} between $60000$ and $80000$ orbits have been 
integrated on a rectangular grid of size $N_x\times N_y\approx 2r_L/\Delta x\times (4/3) r_L/\Delta_y$ of initial 
conditions on the Poincar\'e surface of section with cell side lengths $\Delta x = 0.25$ and $\Delta y= 0.05$.
The grid is centered in the origin of phase space coordinates.
For comparison, for each panel on the right-hand side of Figure \ref{fig:bs2} between $49000$ and $52000$ 
orbits have been integrated on a similar rectangular grid. 
For simplicity, our initial conditions for the calculation of the basins of escape cover both quasiperiodic and chaotic areas
of the corresponding surface of section. We remark that, in principle, it is not necessary to integrate
the orbits in areas with quasiperiodic orbits, since we know that these do not escape.
Also, we do not consider all initial conditions in the same chaotic sea as one and the same orbit. 
We calculated the initial conditions for the velocities in the same way as for the 
Poincar\'e surfaces of section in Section \ref{sec:surf}.


In Figure \ref{fig:bs2}, the red (and yellow) regions correspond to initial conditions for which the escaping
star passes the Lagrangian point $L_1$ (or $L_2$) while it escapes. 
This passing-by condition defines the basins of escape.
We call the corresponding basins of escape, which can be visualised on the surfaces of section,
the $L_1$- (or $L_2$-) basins of escape.
The black regions correspond to initial conditions
where the orbits are trapped and do not escape.
The time of escape $t_e$ for the red and yellow regions is defined as
the time when the escaping star passes the vertical line given by $(x,y) = (\pm r_L, y)$.
As the black regions of Figure \ref{fig:bs2} are concerned, 
we consider an initial condition as non-escaping if its orbit remains
bound for longer than $t_{\rm max}=20000$ (top panels), $t_{\rm max}=50000$ (middle panels) or $t_{\rm max}=100000$ (bottom panels).
The missing symmetry of the basins of escape with respect to the $x$ and $y$ axes are due to the
choice of a subset of orbits with a fixed sign of one component of the velocity vector.
An inspection of Figure \ref{fig:bs2} reveals that for energies close to the critical Jacobi energy 
(and, of course, for energies below the critical Jacobi energy) the allowed phase space volume is coloured
black to a large extent (or nearly totally).

To estimate the size of the chaotic saddle, Figure \ref{fig:backwards} shows 
a backwards integrated basin of escape (top panel) at $e_J =-203.75205349030903$.
We use the same initial conditions for the backwards integrated orbits as for the 
forward integrated ones. The red (and yellow) regions correspond to initial conditions for which the 
escaping star passes $L_1$ (or $L_2$). The black regions correspond to initial conditions for which
the orbit does not escape.
For the backwards integration, the following modifications must be made in the
numerical integration of the orbits:
\begin{enumerate} 
\item The sign of the time step in the Runge-Kutta integrator must be reversed,
\item the sign of the velocities in the equations of motion (\ref{eq:eom}) must be reversed,
\item the sign of the frequencies in the equations of motion (\ref{eq:eom}) must be reversed,
\item  the sign of the velocities in the definition of the surface of section must be reversed.
\end{enumerate}

The chaotic or strange saddle (an invariant chaotic set) 
is given by the intersection of its stable and 
unstable manifolds. 
The stable (unstable) manifolds of the chaotic set
coincide with the boundaries between
the forwards (backwards) integrated $L_1$- and $L_2$-basins of escape 
\citep[c.f.][]{Aguirre2001}.
The stable and unstable manifolds as well as the corresponding basins of escape
are symmetric to each other due to the time symmetry of the equations of motion (\ref{eq:eom}).
In a non-hyperbolic system, there may be tangencies between the stable and unstable
manifolds, i.e. the splitting angle between them may be zero.
A comparison of Figure \ref{fig:bs2} and Figure \ref{fig:backwards} reveals that, in contrast to the
system studied in \citet{Ernst2008},
the phase-spatial extent of the chaotic saddle for the present system is tiny
as compared to the squared Lagrangian radius.
For this reason we did not calculate the chaotic saddle for the system studied in 
the present work.

\begin{figure}
\centering
\includegraphics[height=0.60\textheight]{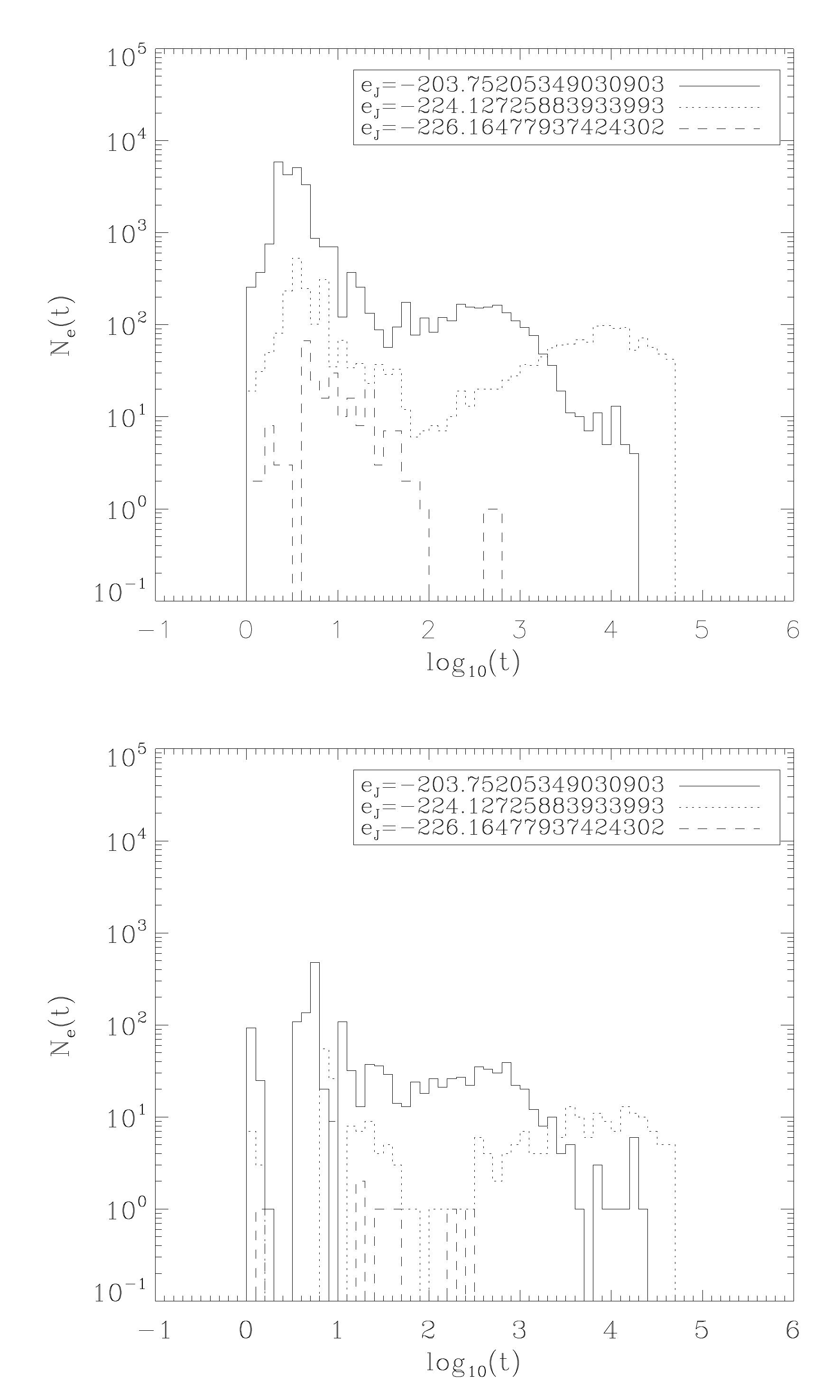} 
\caption{Differential distributions of escape times. 
Top panel: For the $x-y$ basins of escape 
of Figure \ref{fig:bs2}. Bottom panel: For the $x-v_x$ basins of escape of Figure \ref{fig:bs2}.
Solid lines: $e_J =-203.75205349030903$, dotted lines: $e_J =-224.12725883933993$, dashed lines: $e_J =-226.16477937424302$.}
\label{fig:thist3d}
\end{figure}

\begin{figure}
\centering
\includegraphics[height=0.60\textheight]{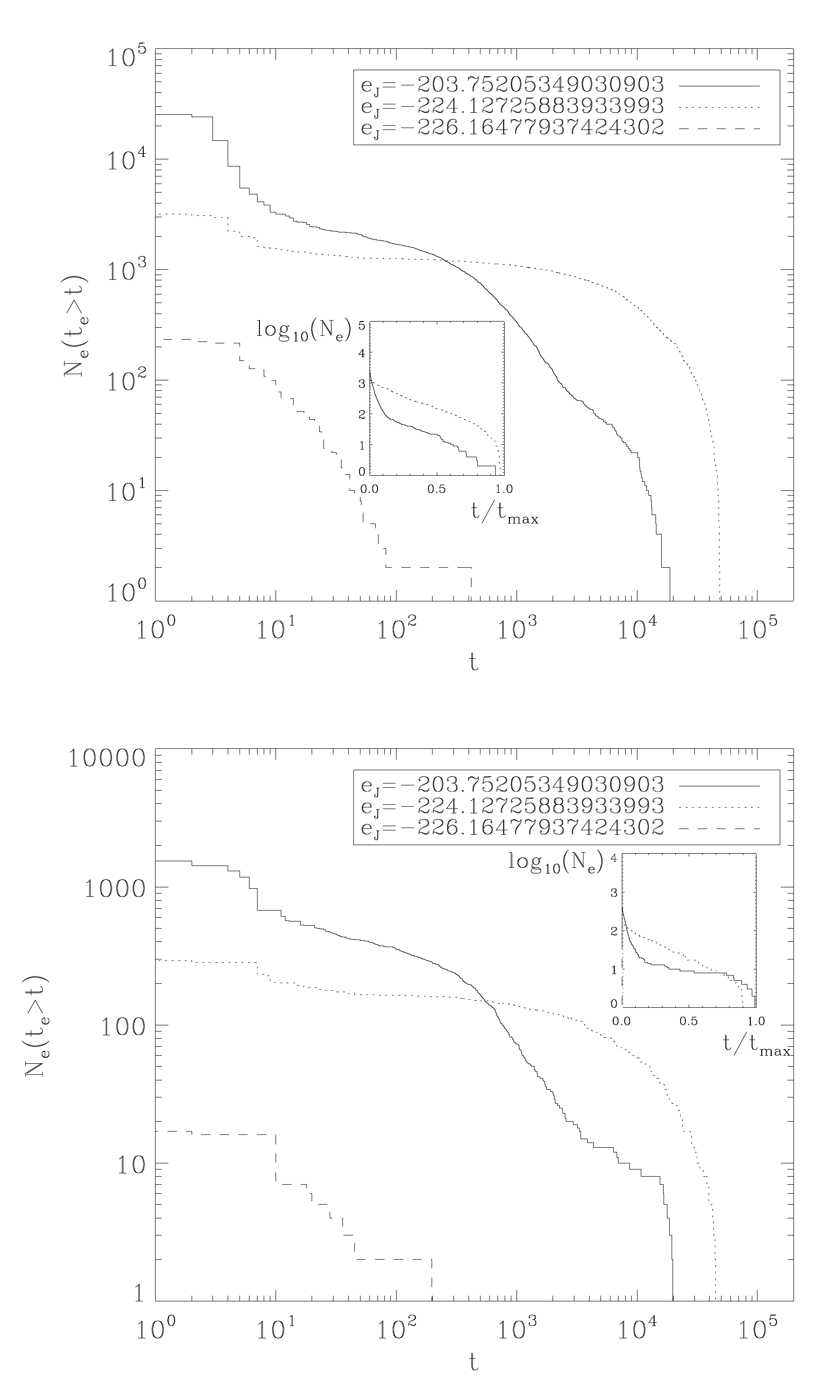} 
\caption{Complementary cumulative distributions of escape times. 
Top panel: For the $x-y$ basins of escape 
of Figure \ref{fig:bs2}. Bottom panel: For the $x-v_x$ basins of escape of Figure \ref{fig:bs2}.
Solid lines: $e_J =-203.75205349030903$, dotted lines: $e_J =-224.12725883933993$, dashed lines: $e_J =-226.16477937424302$. The sharp drop-offs of the solid curves at 
$t\approx 20000 = t_{\rm max}$ and of the dashed curves at $t\approx 50000 = t_{\rm max}$ are due to the artificial cutoffs in the calculation 
program and the fact that only the escaping orbits have been included in the statistics.
The linear-logarithmic insets show the evolution for the two lowest Jacobi energies. 
The histogram for the highest Jacobi energy is not shown in the inset due to the low number of escaping orbits.
The curves in the insets indicate a transition to an exponential decay in the limit of long escape times,
except for the solid curve in the lower panel's inset. }
\label{fig:thist3}
\end{figure}

Figure \ref{fig:tbar} shows the distribution of escape times $t_e$ on surfaces of section 
for the forwards integration at $e_J =-203.75205349030903$ (left panel) and $e_J =-224.12725883933993$ (right panel).
The scale on the color bar is logarithmic. The region within the separatrix, from which orbits do not escape, 
is overplotted with black dots and 
invariant curves on the grey-shaded area.
In particular, Figure \ref{fig:tbar} shows that there are chaotic orbits which do not escape
within time $t_{\rm max}$. 

Figure \ref{fig:thist3d} shows the differential distributions of escape times, i.e. the
escaper number $N_e$ as a function of time $t$. The top panel shows the distributions for 
the $x-y$ basins of escape of Figure \ref{fig:bs2}. 
The bottom panel shows the distributions for the $x-v_x$ basins of 
escape of Figure \ref{fig:bs2}.
The solid lines correspond to $e_J =-203.75205349030903$, the dotted lines correspond to 
$e_J =-224.12725883933993$, and the dashed lines correspond to $e_J =-226.16477937424302$.
The distributions are irregularly shaped,
i.e. they are not monotonically.

In contrast, 
Figure \ref{fig:thist3} shows the complementary cumulative distributions of escape times
on surfaces of section. 
The $x$ axis shows the time $t$, the $y$ axis shows the escaper number with 
escape time $t_e>t$,
which is proportional to the
exceedance or survival probability \citep[c.f.][]{ Altmann2013}. 
As for Figure \ref{fig:thist3d}, the top panel shows the decay for the $x-y$ basins of escape 
of Figure \ref{fig:bs2}. The bottom panel shows the decay for the $x-v_x$ basins of 
escape of Figure \ref{fig:bs2}.
The solid lines correspond to $e_J =-203.75205349030903$, the dotted lines correspond to 
$e_J =-224.12725883933993$, and the dashed lines correspond to $e_J =-226.16477937424302$.
The sharp drop-off of the solid curves at 
$t\approx  t_{\rm max}=20000$ and of the dashed curves at $t\approx t_{\rm max}=50000$ is due to the artificial cutoffs in the calculation 
program and the fact that only the escaping orbits have been included in the statistics. 
For a galaxy which has the size of, say, NGC 1300, the system of units (\ref{eq:units1}) - (\ref{eq:units3})
yields that these maximum times correspond to $2000$ and $5000$ Gyr, respectively, both of which are 
well above $100$ Hubble times.
The linear-logarithmic insets show the evolution for the two lowest Jacobi energies. 
The histogram for the highest Jacobi energy is not shown in the inset due to the low number of escaping orbits.
The curves in the insets, except for the solid curve in the lower panel's inset, 
indicate a transition to an exponential decay law in the limit of 
long escape times.  The slope of the exponential 
decay law characterizes the strange chaotic saddle \citep{Kadanoff1984}. It can be seen that the 
solid curve in the lower panel's inset is, in contrast to the other curves in the insets, not straight in the 
shown range. The reason may be that for this parameter set the chaotic saddle vanishes.

\subsection{Spiral arms}

\begin{figure*}
\includegraphics[height=0.35\textheight]{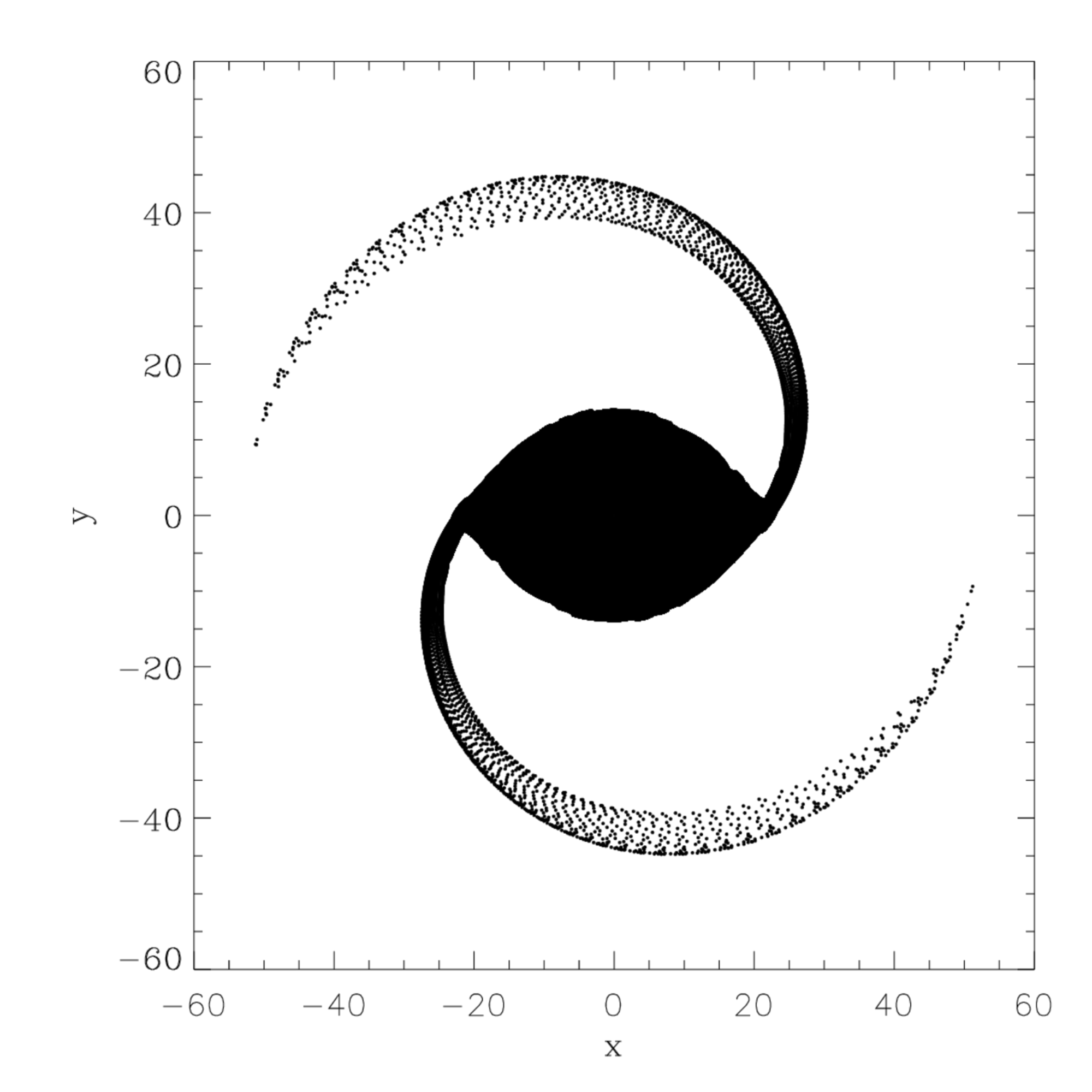} 
\includegraphics[height=0.35\textheight]{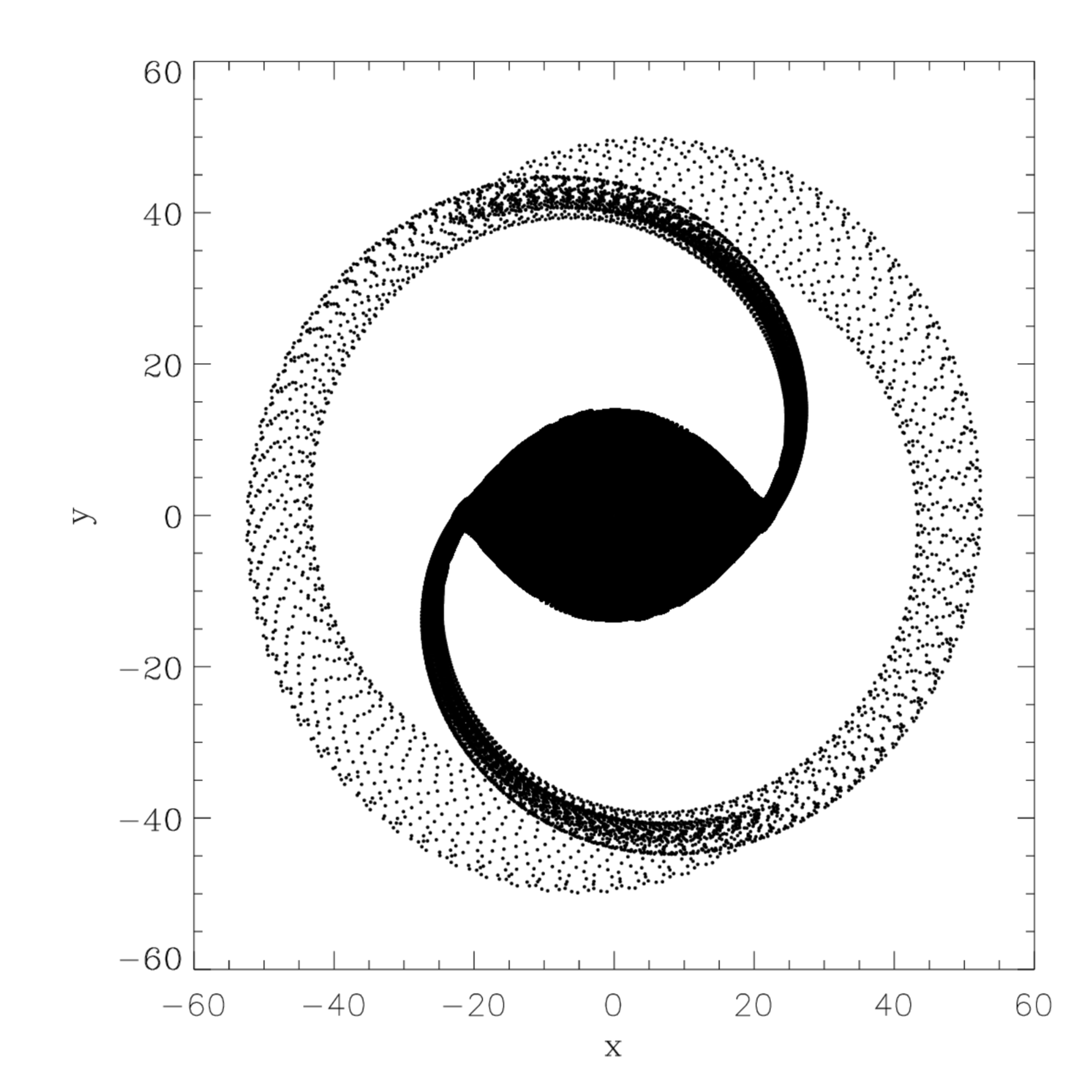} 
\caption{Formation of spiral arms and a ring at $e_J =-224.12725883933993$ 
in the effective potential of Eqn. (\ref{eq:zotospotential}).
Left: At $t=7.5$. Spiral arms have formed. Right: At $t=10.0$. A ring has formed.}
\label{fig:spirals}
\end{figure*}

In the case of star clusters, escaping orbits form so-called
tidal tails \citep{Odenkirchen2003,DiMatteo2005,Kuepper2008,Just2009,Ernst2009,Kuepper2010,Berentzen2012}. 
Based on the similarity of the correspondig effective potentials, we would like to 
point out here the striking similarity between spiral arms of barred spiral galaxies 
and star cluster tidal tails 
and mention the hypothesis that the spiral arms of barred 
spiral galaxies such as NGC 1300 are the equivalent of such star cluster tidal tails forming
in an effective potential similar to that of Eqn. (\ref{eq:zotospotential}) below 
which has two saddle points
(see Figure \ref{fig:effpotbar2} for the case of a two-armed
barred spiral galaxy).
Since star cluster tidal tails follow the curvature of the
orbit of the star cluster around the galactic centre, they are nearly straight for orbits with large 
galactocentric distances \citep[see Figure 1 in][with the effective potential shown in their Figure 2]{Just2009}. On the other hand, the spiral arm ``tidal tails'' of barred spiral galaxies are not caused by the tidal field but by the non-axisymmetric bar-like
perturbation, wind up around the banana-shaped isolines of the effective potential in which $L_4/L_5$ are enclosed
and form the spiral-like shape which is prominent in 
barred spirals with two spiral arms which emerge from the ends of the bar
\citep[][Plate 10 showing NGC 1300]{Binney2008}.

In Figure \ref{fig:spirals} we show at $e_J =-224.12725883933993$ that this scenario is viable. 
We have modified our integration program for the calculation of the basins of escape to yield output
of all orbit trajectories for a three-dimensional grid of size $N_x\times N_y \times N_{v_x}\approx 11 \times 800 \times 3$  with $\Delta x = 4.0, \Delta y = 0.025$ and $\Delta v_x = 5.0$. The grid is centred in the origin of 
phase space coordinates.
We allow for both signs of $v_y$.
The snapshot in the left-hand panel of Figure \ref{fig:spirals}
shows that at $t=7.5$, which corresponds to 750 Myrs in the units given by Eqns. (\ref{eq:units1}) - (\ref{eq:units3}), 
two spiral arms with a similar morphology to those in \citet[][his Figure 1]{Patsis2006} have formed. 
The snapshot in the right-hand panel of Figure \ref{fig:spirals}
shows the situation at $t=10.0$ (1 Gyr), where a ring has formed. The standard orbit integration routine,
which is used to obtain Figure \ref{fig:orbits2d},
plots a point at every integration step. 
However, in Figure \ref{fig:spirals}, the density of points along one stellar orbit is taken to be proportional
to the velocity of a particle. In particular, a point is plotted, if an integer counter variable, which is increased
by one in every integration step, exceeds the velocity 
of the particle. 
In such a way we simulate a real $N$-body simulation of such a system: The density of particles
will be highest where the velocity is lowest. Similarly, clumps form in tidal tails of star clusters at positions where the
velocity of the escaping particles is lowest \citep{Kuepper2008}.

The two morphologies in Figure \ref{fig:spirals} are also discussed in the papers by \citet{Athanassoula2009, Athanassoula2009b}. These papers provide the connection between the bar strength and the corresponding
morphological types. For example, they predict that, if the non-axisymmetric forcing is relatively low
the resulting morphology will be an $R_1$ or $R_1'$ ring while if it is stronger, it will be a spiral or one of the remaining types of rings ($R_2, R_1R_2$ etc.). In our case, we see
that the evolutionary state of the barred spiral galaxy is of relevance as well. Rings may be more evolved than spirals \citep[cf.][]{Athanassoula2012}.



\section{Discussion}

\label{sec:discussion}

We have studied the region in the close vicinity of and inside the critical volume of a 
galactic potential with a bar. The critical volume is defined as the volume which 
encloses the last closed equipotential surface of the effective potential.
The present paper is complementary to the series of papers by \citet{Athanassoula2009, Athanassoula2009b, Athanassoula2010, Athanassoula2012}. While the latter papers are concerned with the morphology,
application of the manifold theory to and comparison with real galaxies,
the present paper gives quantitative information on the escapes.
We have particularly studied the physics in the barred four-component 
effective potential in Eqn. (\ref{eq:zotospotential}) by \cite{Zotos2012}.
We have calculated Poincar\'e surfaces of section and the basins of escape at different
values of the Jacobi energy, a few examples of typical orbits,
the distribution of escape times on a surface of section, differential and 
complementary cumulative
distributions of escape times. We have also studied the behaviour of
the escaped particles outside of the Lagrangian radius, where they form spiral arms or a ring.

\medskip

We state the main conclusions of this work, 
which are valid for the system given by Eqn. (\ref{eq:zotospotential}) with our corresponding choice of the
parameters,  as follows:

\begin{enumerate}
\item We have found numerical evidence for the existence of a 
separatrix in phase space 
in both the Poincar\'e surfaces of section and the basins of escape
which hinders particles from escaping out of the bar region. 
While the adelphic integral hinders quasiperiodic orbits from escaping,
the separatrix prevents chaotic orbits from escaping. We have found that there are chaotic
orbits which do not escape within $t_{\rm max}$.
\item
\begin{enumerate} 
\item The late-time exponential decay related to 
the chaotic saddle is not relevant for a NGC 1300 sized barred galaxy
since the corresponding escape times are well above a Hubble time. 
However, the early-time escape process within a Hubble time is relevant (see Figure \ref{fig:thist3d}).
Note that these two statements still hold for a galaxy which has one tenth 
the size in both $x$ and $y$ directions and one hundredth of the mass of NGC 1300.
\item The phase-spatial extent of the 
chaotic saddle is tiny as compared with the squared Lagrangian radius.
\end{enumerate}
\item We have presented evidence for a striking similarity of spiral arms of barred 
spiral galaxies to tidal tails of star clusters. 
\end{enumerate}

Future research may put the above-mentioned conclusion (i), (ii) (a) and (b), which hold for the
special case of the potential of Eqn. (\ref{eq:zotospotential}), on firmer grounds 
with more numerical evidence for other cases with modified bar potentials as those used in
\citet{Athanassoula2009, Athanassoula2009b, Athanassoula2010} and identify them
as generally valid in typical barred galactic potentials or falsify them for 
the general case of galactic potentials with bars. 

Concerning conclusion (iii), if it is true that the spiral arms of barred spiral galaxies are formed 
out of escaping particles from the bar region
we may think of the formation scenario of barred spiral galaxies as follows \citep[cf.][]{Athanassoula2012}:

\begin{enumerate}
\item An axisymmetric rotating stellar system forms.
\item At a certain redshift the system gets unstable to the 
formation of a bar (i.e. it develops a non-axisymmetric bar-like 
perturbation).
\item Escaping particles with Jacobi energies higher than the critical one
form spiral arms (similar to the formation of tidal tails in star clusters). 
At the same time the bar strength decreases as witnessed in  \citet{Athanassoula2012}.
\item The barred spiral galaxy has formed.
\end{enumerate}

Future $N$-body simulations may elaborate the different shapes and morphologies of barred spiral galaxies
\citep[e.g.][]{Buta2011} and measure the amplitudes of the spiral perturbations.
Moreover,  results from the theory of the dissolution of star clusters may be
applied to the problem of the formation of barred spiral galaxies. Also, the manifold flux-tube theory
of Athanassoula et al. may be applied to the formation of tidal tails in star clusters.

\section{Acknowledgements}

Both authors would like to thank 
Dr. Thorsten Lisker for pointing out a few references to them and the 
anonymous referee for thoughtful comments on the manuscript which greatly helped to improve the paper.
AE acknowledges partial financial support by grant JU 404/3-1 of the Deutsche 
Forschungsgemeinschaft (DFG)
and hospitality and partial financial support through Silk Road Project at National Astronomical 
Observatories of Chinese Academy of Sciences (NAOC) in Beijing, China, Grant Number 2009S1-5.
TP acknowledges financial support through a Forschungskredit of the University of 
Z\"urich, grant no. FK-13-112. The APOD website (http://apod.nasa.gov) has been used
as a further source of information.
\bibliographystyle{mn2e}
\bibliography{chaos}

\appendix

\bsp

\label{lastpage}

\end{document}